\documentclass[a4paper]{jpconf}

\usepackage{graphicx}  % needed for figures
\usepackage{dcolumn}   % needed for some tables
\usepackage{bm}        % for math
\usepackage{amsfonts,amsmath,amssymb,mathrsfs}
\usepackage{color}

\newcommand{\eq}{\begin{equation}}
\newcommand{\eeq}{\end{equation}}
\newcommand{\be}{\begin{equation}}
\newcommand{\ee}{\end{equation}}
\newcommand{\bea}{\begin{eqnarray}}

\newcommand{\eea}{\end{eqnarray}}
\newcommand{\vta}[1]{\vert \boldsymbol{a}_{#1}        \vert}

\newcommand{\vtl}   {\vert \boldsymbol{       {\ell}} \vert}

\newcommand{\ie}{\textit{i.e.,}~}

\newcommand{\prd}{Phys. Rev. D}
\newcommand{\apj}{Astrophys. J.}
\newcommand{\apjl}{Astrophys. J. Lett.}
\newcommand{\prl}{Phys. Rev. Lett.}

\newcommand{\mnras}{Mon. Not. Roy. Astron. Soc.}

\begin{document}
\title{The importance of precession in modelling the direction of the final spin from a black-hole merger}

\author{Enrico Barausse}

\address{Center for Fundamental Physics, University of Maryland, College Park, MD 20742-4111, USA}
% \ead{barausse@umd.edu}

\begin{abstract}
The prediction of the spin of the black hole resulting from
the merger of a generic black-hole binary system is of great
importance to study the cosmological evolution of 
supermassive black holes. Several attempts have been recently made
to model the spin via simple expressions
exploiting the results of numerical-relativity simulations.
Here, I first review the derivation of a formula, proposed in Ref.~\cite{new}, which
accurately predicts the final spin magnitude and direction when applied to binaries with separations of hundred or thousands
of gravitational radii.  This makes my formula particularly suitable for
cosmological merger-trees and N-body simulations, which provide the spins and angular momentum of the two
black holes when their separation is of thousands of gravitational radii. More importantly, I investigate the physical 
reason behind the good agreement between my formula and numerical relativity simulations, and nail it down
to the fact that my formula takes into account the post-Newtonian precession of the spins and angular momentum in
a consistent manner.
\end{abstract}

% \section*{Introduction}
The dynamics of black-hole (BH) binaries is a very complex problem which has been solved 
only very recently through time-expensive
numerical-relativity (NR) calculations. 
However, it has been shown that the dimensionless spin of the remnant from a BH binary merger,
$\boldsymbol{a}_{\rm fin}=\boldsymbol{S}_{\rm fin}/M_{\rm fin}^2$, can
be described with simple prescriptions based
on point particles~\cite{scott,bkl,kesden}, on fits to the NR
data~\cite{fit1,fit2,faunew,boyle1,boyle2}, or on a combination of
the two approaches~\cite{old}. 
These formulas are useful because they
provide information over the entire 7-dimensional space of
parameters for BH binaries in quasi-circular orbits, namely: the 
mass ratio $q\equiv M_2/M_1$ and the six components of the initial
dimensionless spin vectors
$\boldsymbol{a}_{1,2}=\boldsymbol{S}_{1,2}/M_{1,2}^2$. Such parameter
space could in principle be investigated entirely via NR calculations;
in practice, however, the simulations are very expensive and
restricted to $q=0.1$--$1$. Also, these formulas have 
applications in astrophysics, where they could provide
information on the properties of massive-star binary systems; in cosmology, where supermassive BHs (SMBHs) are
believed to assemble through accretion and
mergers; in gravitational-wave astronomy, where
the \textit{a priori} knowledge of the final spin can 
help the detection.

While the different expressions for the spin norm,
$|\boldsymbol{a}_{\rm fin}|$, are in good agreement with the results of NR simulations, the predictions for the final spin direction, 
$\hat{\boldsymbol{a}}_{\rm fin}\equiv {\boldsymbol{a}}_{\rm fin}/|{\boldsymbol{a}}_{\rm fin}|$, do not agree well with one another 
and are all essentially \textit{imprecise} when the binaries are widely separated. This is
because all expressions are built from and model the typical NR
binaries and hence the dynamics of the last few orbits before the
merger. Because it does not account systematically for the precession
of the orbital angular momentum $\boldsymbol{L}$, the prediction for
$\hat{\boldsymbol{a}}_{\rm fin}$ depends on the separation of the
binary and is therefore of little use for applications, such as
cosmological merger-trees or N-body simulations, that provide the spins of the two BHs at separations of thousands of gravitational
radii. Although one
could use the PN equations to evolve a widely-separated binary 
to a separation of few gravitational radii and then apply
the formulas, this makes the formulas
difficult to use and implement. In Ref.~\cite{new} we followed instead a different approach and showed that it is possible
to derive a formula for the final spin that takes into account the precession of the spins and that is, therefore,
applicable to binaries with arbitrary separations. 

% \section*{The formula}

The derivation of the formula, for more details about which we refer the reader to
Ref.~\cite{new}, is based on the following assumptions:

\smallskip

\noindent\textit{(i) When the spins are parallel to $\boldsymbol{L}$ (either aligned or antialigned), 
the formula must reduce to the fit of the NR results
presented in Refs.~\cite{new,old}, namely} 
\begin{equation}
\label{eqspin_uneqmass}
a_{\rm fin}=\tilde{a}
+\tilde{a} \nu (s_{4}\tilde{a}+s_{5}\nu + t_0)
+ \nu(2\sqrt{3}+t_2\nu+t_{3}\nu^2)\,,
\end{equation}
where $\nu \equiv M_1M_2/(M_1+M_2)^2$ is the symmetric mass ratio,
$\tilde{a}\equiv(a_1 + a_2 q^2)/(1+q^2)$, and\footnote{The number of free parameters of the fit is actually \textit{four}, since
 $t_0$, $t_2$, $t_3$, $s_4$ and $s_5$ must satisfy the constraint
\begin{equation*}
 \label{eq:constraint}
 a_{\rm fin}= \frac{\sqrt{3}}{2} + \frac{t_2}{16} + \frac{t_3}{64} 
 = 0.68646 \pm 0.00004\,.
\end{equation*}
which follows from the results obtained by Ref.~\cite{caltech_cornell} for
equal-mass non-spinning BHs.}
\begin{align}
\label{eq:coeff}
&s_4 = -0.1229\pm0.0075\,,\quad s_5 = 0.4537\pm0.1463\,,\nonumber\\ 
&t_0=-2.8904\pm0.0359\,,\quad t_2=-3.5171 \pm 0.1208\,,\quad t_3 = 2.5763\pm0.4833\,.
\end{align}

\smallskip
\noindent\textit{(ii) The mass $M_{\rm rad}$ radiated to gravitational waves can be neglected} 
\ie $M_{\rm fin} = M \equiv M_1 + M_2$. 
The reason why assumption \textit{(i)} is
reasonable here is that $M_{\rm rad}$ is largest for aligned
binaries but these are also the ones fitted by expression~\eqref{eqspin_uneqmass}. In this way, the mass losses are
automatically accounted for by the values of the coefficients $t_0$, $t_2$, $t_3$, $s_4$ and $s_5$.

\smallskip
\noindent\textit{(iii) The  norms $|\boldsymbol{S}_1|$,
  $|\boldsymbol{S}_2|$, $|\boldsymbol{\tilde{\ell}}|$ do not depend on
  the separation of the binary $r$}, with $\boldsymbol{\tilde{\ell}}$ being %defined as 
\begin{equation}
\label{ell_def}
\boldsymbol{\tilde{\ell}}(r)\equiv \boldsymbol{S}_{\rm
  fin}-[\boldsymbol{S}_1(r)+\boldsymbol{S}_2(r)]=
\boldsymbol{L}(r)-\boldsymbol{J}_{\rm rad}(r)\,, 
\end{equation} 
where $\boldsymbol{S}_1(r)$, $\boldsymbol{S}_2(r)$ and
$\boldsymbol{L}(r)$ are the spins and the orbital angular momentum
at the separation $r$ and $\boldsymbol{J}_{\rm
  rad}(r)$ is the angular momentum radiated from $r$ to the
merger. While the constancy of $|\boldsymbol{S}_1|$ and $|\boldsymbol{S}_2|$ is a very good
assumption for BHs,
the constancy of
 $|\boldsymbol{\tilde{\ell}}|$ is heuristic and based on the
idea that the merger takes place at an ``effective''
innermost stable circular orbit (ISCO), so that $|\boldsymbol{\tilde{\ell}}|$ can be interpreted
as the residual orbital angular momentum contributing
to $\boldsymbol{S}_{\rm fin}$.

\smallskip 
\noindent\textit{(iv) The final spin $\boldsymbol{S}_{\rm fin}$ is
  parallel to the initial total angular momentum
  $\boldsymbol{J}(r_{\rm in})\equiv\boldsymbol{S}_1(r_{\rm in})+
  \boldsymbol{S}_2(r_{\rm in})+\boldsymbol{L}(r_{\rm in})$.} 
% This is motivated by the PN approximation. It has been in fact 
% shown by~\cite{apostolatos} that within the adiabatic approximation
% the secular angular-momentum losses via gravitational radiation are along
% $\boldsymbol{J}$. This is because as $\boldsymbol{L}$ rotates around
% $\boldsymbol{J}$, the emission orthogonal to $\boldsymbol{J}$ averages
% out.

\smallskip
\noindent\textit{(v) The angle between $\boldsymbol{L}$ and
  $\boldsymbol{S}\equiv\boldsymbol{S}_1+ \boldsymbol{S}_2$ and the
  angle between $\boldsymbol{S}_1$ and $\boldsymbol{S}_1$ are constant
  during the inspiral, although $\boldsymbol{L}$ and $\boldsymbol{S}$
  precess around $\boldsymbol{J}$.}

\smallskip

Assumptions  $\textit{(iv)}$ and
$\textit{(v)}$ are motivated by the PN approximation. It has been in fact 
shown by Ref.~\cite{apostolatos} that within the adiabatic approximation
the secular angular-momentum losses via gravitational radiation are along
$\boldsymbol{J}$. This is because as $\boldsymbol{L}$ rotates around
$\boldsymbol{J}$, the emission orthogonal to $\boldsymbol{J}$ averages out. Therefore, 
$\boldsymbol{S}$ and $\boldsymbol{L}$
essentially precess around the direction $\boldsymbol{\hat{J}}$, which remains nearly constant (\textit{cf.} the detailed
  discussion in Ref.~\cite{apostolatos}), and the angle between $\boldsymbol{L}$ and $\boldsymbol{S}$ and that between the two
spins remain constant as well.

\smallskip
\noindent\textit{(vi) When the initial spin vectors are equal and
  opposite and the masses are equal, the spin of the final BH is the
  same as for nonspinning binaries}.
 Besides being physically reasonable -- if the spins are equal and opposite, their
contributions are expected to cancel out -- this assumption is confirmed by 
NR simulations and by the leading-order PN spin-spin and spin-orbit couplings. 

\smallskip

These assumptions are sufficient to derive an expression for both the magnitude and the direction of the final spin~\cite{new}. In particular,
the final spin norm is given by 
 \begin{eqnarray}
 \label{eq:general}
 &\vert \boldsymbol{a}_{\rm fin}\vert=
 \frac{1}{(1+q)^2}\Big[ \vta{1}^2 + \vta{2}^2 q^4+
  2 {\vert \boldsymbol{a}_2\vert}{\vert 
 \boldsymbol{a}_1\vert} q^2 \cos \alpha\,+
 \nonumber\\ 
 & \hskip 0.2cm
 2\left(
      {\vert \boldsymbol{a}_1\vert}\cos \beta +
      {\vert \boldsymbol{a}_2\vert} q^2  \cos \gamma
 \right) {\vert \boldsymbol{{\ell}} \vert}{q}+\vert \boldsymbol{{\ell}}\vert^2 q^2
 \Big]^{1/2},
 \end{eqnarray}
 where 
\begin{eqnarray}
\label{eq:L2}
&&\vtl = 2 \sqrt{3}+ t_2 \nu + t_3 \nu^2 +
 \frac{s_4}{(1+q^2)^2} \left(\vta{1}^2 + \vta{2}^2 q^4 
	+ 2 \vta{1} \vta{2} q^2 \cos\alpha)\right) + 
\nonumber \\
&&
\left(\frac{s_5 \nu + t_0 + 2}{1+q^2}\right)
	\left(\vta{1}\cos{\beta} + 
        \vta{2} q^2 \cos{\gamma}\right)\,,
\nonumber 
\end{eqnarray}
and the angles $\alpha, \beta$ and $\gamma$ are defined by
 \begin{equation}
 \label{cosines}
 \cos \alpha \equiv
 {\boldsymbol{\hat{a}}_1(r_{\rm in})\cdot\boldsymbol{\hat{a}}_2}(r_{\rm in})
 \,,
 \hskip 0.3cm
 \cos \beta \equiv
  \boldsymbol{\hat a}_1(r_{\rm in})\cdot\boldsymbol{\hat{{L}}}(r_{\rm in})\,,
 \hskip 0.3cm
 \cos \gamma \equiv
 \boldsymbol{\hat{a}}_2(r_{\rm in})\cdot\boldsymbol{\hat{{L}}}(r_{\rm in})\,.
 \end{equation}
The angle $\theta_{\rm fin}$ between the
final spin and the initial orbital angular momentum
$\boldsymbol{L}(r_{\rm in})$ is instead
\begin{equation}
\label{eq:angle} \cos \theta_{\rm fin} =
\boldsymbol{\hat{L}}(r_{\rm in}) \cdot \boldsymbol{\hat{J}}(r_{\rm
  in})\,.  
\end{equation}

\begin{figure}
  \begin{center}
    \begin{tabular}{cc}
      \resizebox{80mm}{!}{\includegraphics{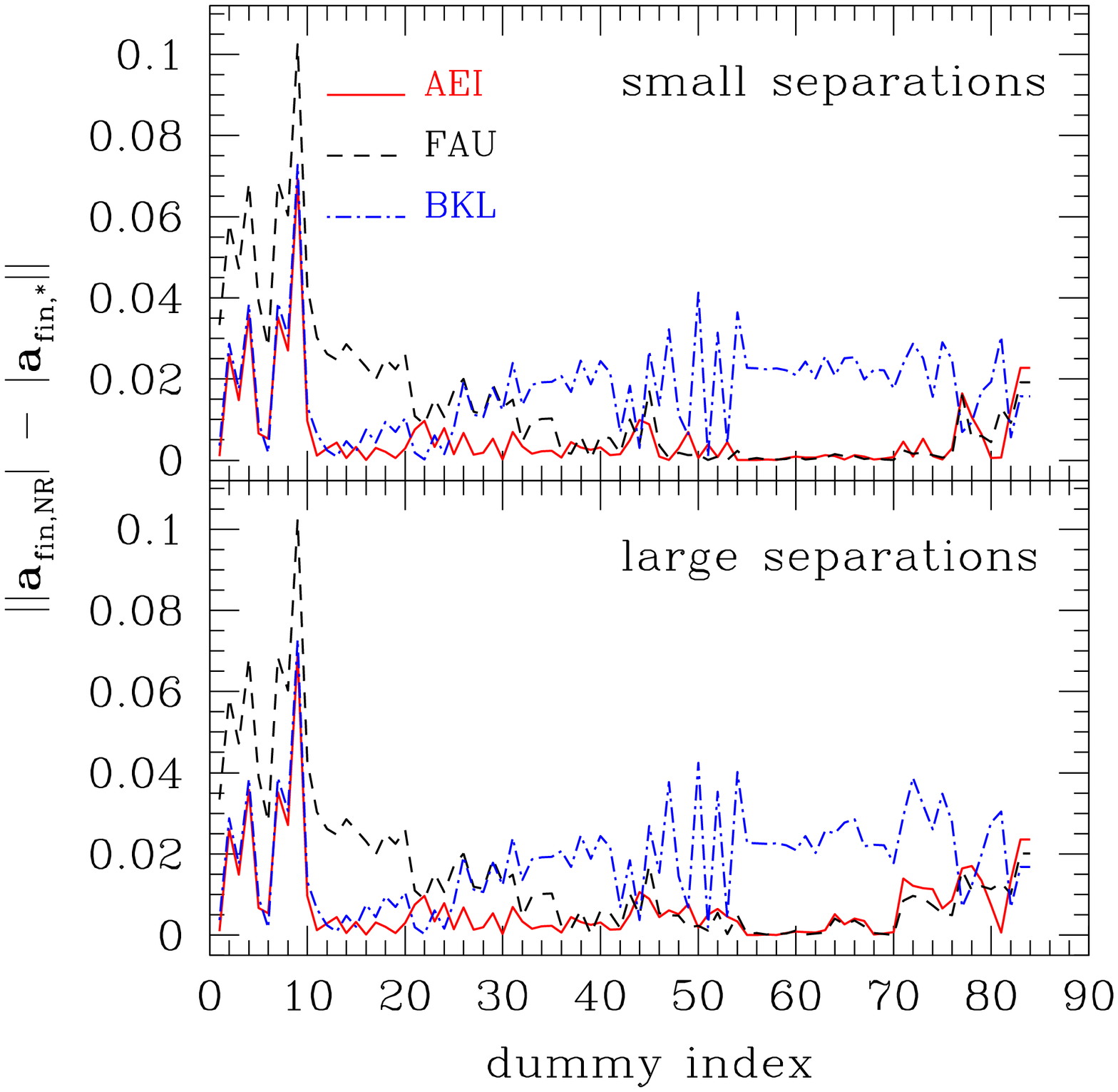}} &
      \resizebox{80mm}{!}{\includegraphics{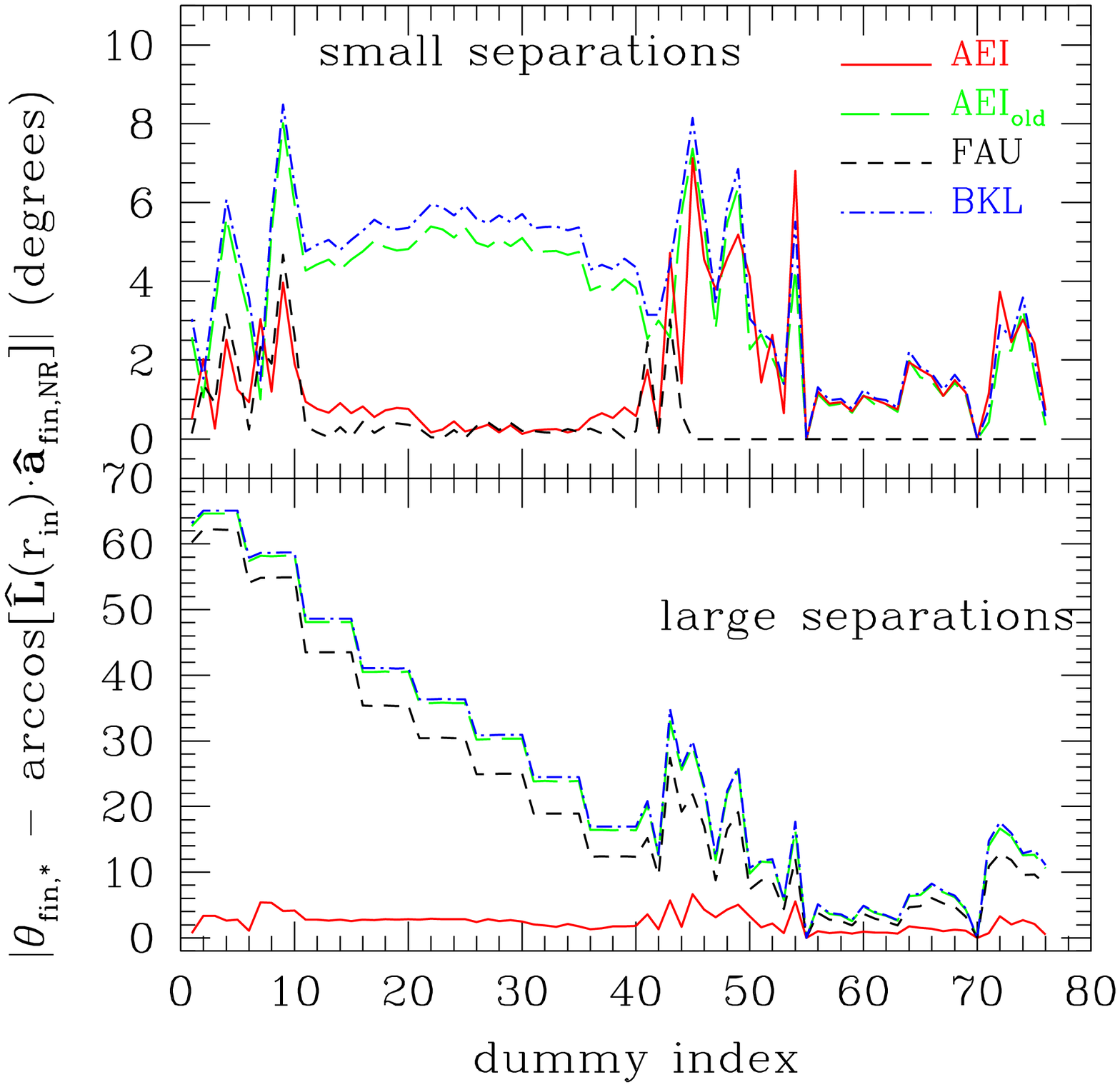}} \\
    \end{tabular}
    \caption{\small \textit{Left:} The upper panel shows the error
  $\vert\vert\boldsymbol{a}_{\rm fin,
    NR}\vert-\vert\boldsymbol{a}_{\rm fin, *}\vert\vert$ (``*'' being
  either ``AEI'', ``FAU'' or ``BKL'') of the various formulas for the
  final spin norm, when applied to the
  \textit{small-separation} configurations corresponding to the
  initial data of the simulations of Refs.~\cite{RITnew} (indices
  1-40),~\cite{sp34} (indices 41-42),~\cite{sp6} (index
  43),~\cite{faunew} (indices 44-76) and~\cite{fau} (indices 77-84). The lower panel shows instead the
  \textit{maximum} error $\vert\vert\boldsymbol{a}_{\rm fin,
    NR}\vert-\vert\boldsymbol{a}_{\rm fin, *}\vert\vert$ when the
  configurations are evolved back in time up to \textit{large
    separations} of $r= 2\times 10^4 M$. Although the new AEI expression is slightly better, all
  formulas give accurate predictions for
  $\vert\boldsymbol{a}_{\rm fin}\vert$, both for small and large
  separations.
 Note that the larger errors
 for the indices 1--10 are due to the larger truncation errors
 affecting those simulations (see text for details).
\textit{Right:} The same as in the left panel but
  for the inclination angle error
  $\vert\theta_{\rm fin, *}-\arccos{[\boldsymbol{\hat{L}}(r_{\rm in}) \cdot \boldsymbol{\hat{a}}_{\rm
  fin,NR}]}\vert$ and without the data of Ref.~\cite{fau}, for which the
  final spin direction has not been published. 
  The new AEI expression is accurate both
  for small and large separations, while the other ones become
  imprecise for large separations.
}
\label{fig:combo}
  \end{center}
\end{figure}

%  \section*{Comparison with numerical relativity data}

We will now test our expressions~\eqref{eq:general} and~\eqref{eq:L2}
for $|\boldsymbol{a}_{\rm fin}|$ and our expression~\eqref{eq:angle} for $\theta_{\rm fin}$ 
against the NR simulations for \textit{generic} binaries (\ie
with spins \textit{not} parallel to $\boldsymbol{L}$) published so
far.
%  \cite{fau,sp34,sp6,RITnew,faunew}. 
Also, we will compare our 
predictions (AEI) with those of similar formulas suggested by Refs.~\cite{bkl} (BKL),~\cite{old} (AEI old)
and~\cite{faunew} (FAU). The comparison consists of two
steps. First, we compare the different predictions using as
input the initial data of the NR simulations, in which the binaries have 
\textit{small separations} ($r_{\rm in}\lesssim 10\, M$). 
Second, using binaries at \textit{large separations} ($r_{\rm in} \leq
2 \times 10^4\, M$), for which the dynamics starts being dominated by gravitational-wave emission and thus
of direct relevance for cosmological investigations. More precisely, 
we evolve the NR initial configurations back in time up to a separation of $2 \times 10^4\, M$ using the PN
equations in the quasi-circular limit\footnote{In particular, following Ref.~\cite{bcv}, 
we use the precession equations for the spins and the
angular momentum at 2 PN order and the rate of change of the frequency at 3.5 PN order (with spin terms
included up to 2 PN order).}, calculating the predictions of
the different formulas at each step and considering the maximum error for each formula.
(The results that we present, however, do not change significantly
if we integrate only up to a separation of $\sim 200 M$.)

The upper left panel of Fig.~\ref{fig:combo} shows the predictions of the various formulas
for $\vert\boldsymbol{a}_{\rm fin}\vert$, when applied to the small-separation
configurations corresponding to the initial data of the NR simulations
(see caption for details). In particular, it reports the
error $\vert\vert\boldsymbol{a}_{\rm fin,
  NR}\vert-\vert\boldsymbol{a}_{\rm fin, *}\vert\vert$, where ``*''
stands either for ``AEI'' (which, as mentioned in Ref.~\cite{new}, 
gives the same predictions for $\vert\boldsymbol{a}_{\rm fin}\vert$ 
as ``AEI old''), ``FAU'' or ``BKL''.
The lower left panel shows instead the \textit{maximum}
error when the configurations are evolved back in time up to $r_{\rm in}= 2\times 10^4 M$. 
Although the AEI expression is slightly better, all the formulas give accurate predictions
for the final spin \textit{norm}, both for small and large
separations.  Note that the larger errors
for the indices 1--10, which correspond to the  simulations of Ref.~\cite{RITnew} 
with small mass ratios ($q=0.13$--$0.17$), are most likely due to the larger truncation errors
affecting those simulations (see Ref.~\cite{RITnew}, sec. IIIA).

The situation is very different when considering the final spin
\textit{direction}. In particular, 
the right panels of Fig.~\ref{fig:combo} report the inclination-angle error, $\vert\theta_{\rm fin, *}-\arccos{[\boldsymbol{\hat{L}}(r_{\rm in}) \cdot \boldsymbol{\hat{a}}_{\rm
  fin,NR}]}\vert$,
for  the data in the left panels, but for those of Ref.~\cite{fau}, 
for which the final spin direction was not
published. Clearly, when considering small-separation binaries (upper right panel), our new formula
performs slightly better than the ``BKL'' and ``AEI
old'' formulas, but it is not better than the ``FAU'' formula. Indeed,
the latter is exact by construction for the indices
34--66. This is because for such data the final spin direction
has not been published and it has been here reconstructed
using the FAU formula applied to the configurations of Table II
in Ref.~\cite{faunew}. However, when considering 
large-separation binaries (lower right panel), our new formula clearly performs much better
than all the other ones. For instance, our error for
$\theta_{\rm fin}$ is below 7 degrees, for \textit{any} separation, %$r_{\rm in}\leq 2\times 10^4 M$,
while it can be as large as 70 degrees with the older formulas. (The
``steps'' in the lower right panel 
reflect the different sequences of Table I
of Ref.~\cite{RITnew}.) 

\begin{figure}
  \begin{center}
    \begin{tabular}{cc}
      \resizebox{78mm}{!}{\includegraphics{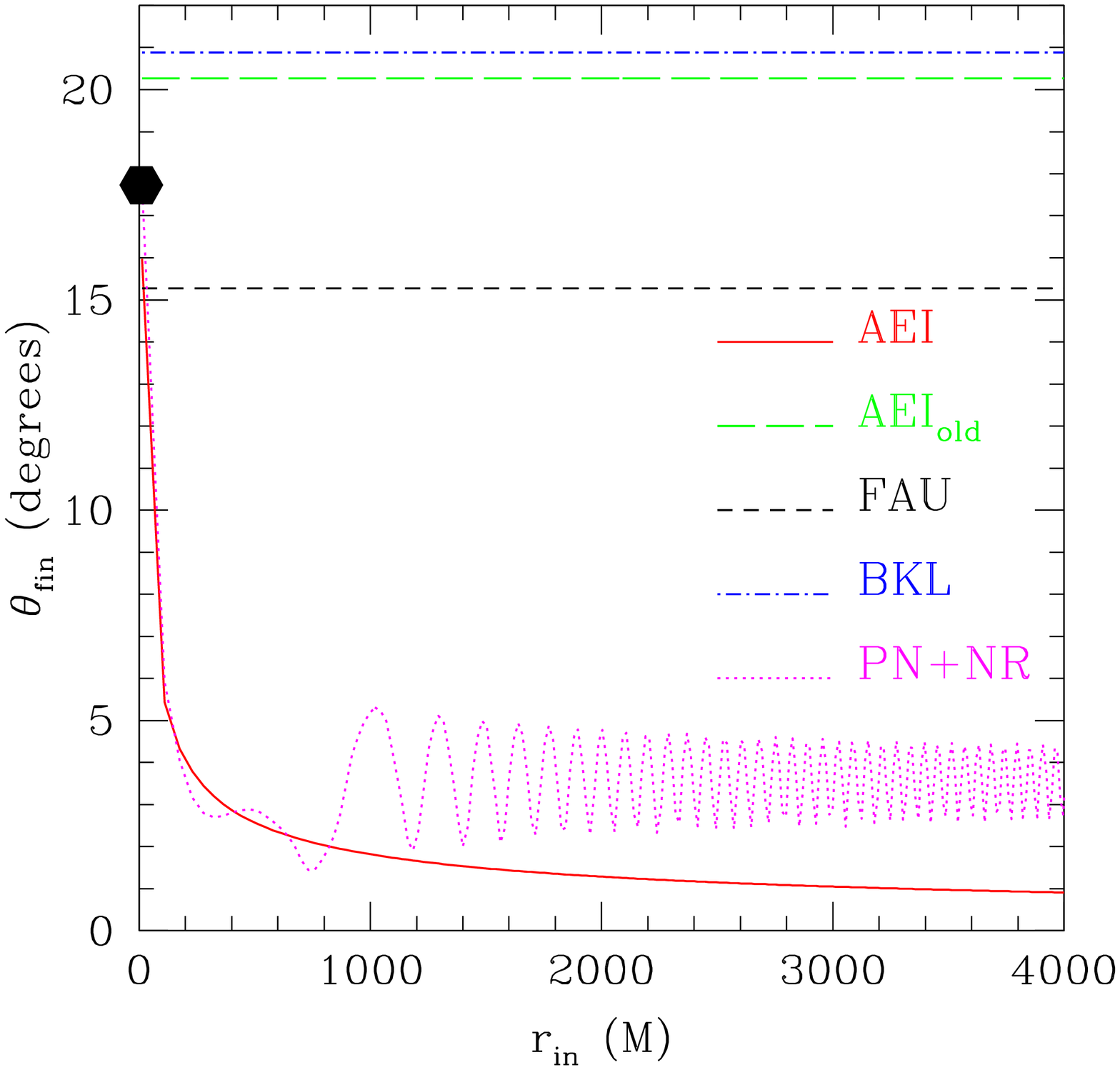}} &
      \resizebox{78mm}{!}{\includegraphics{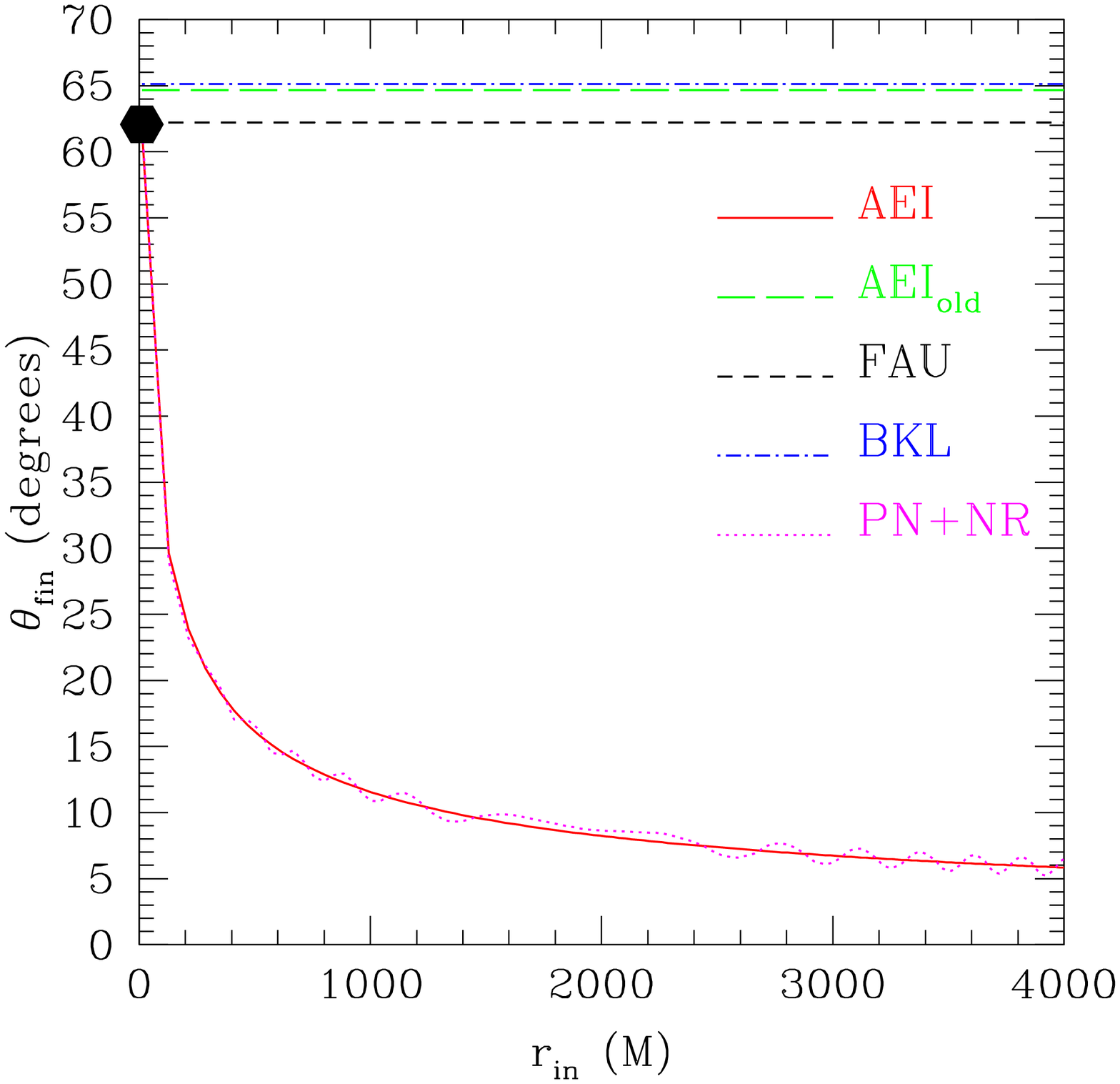}} \\
    \end{tabular}
\caption{\small \textit{Left:} Predictions of the different formulas for $\theta_{\rm
    fin}$ for the binary ``SP3'' of Ref.~\cite{sp34} as a function
  of the initial separation $r_{\rm in}$. Shown with the ``NR+PN''
  curve is the angle $\arccos{[\boldsymbol{\hat{L}}(r_{\rm in}) \cdot
      \boldsymbol{\hat{a}}_{\rm fin,NR}]}$, where
  $\boldsymbol{\hat{L}}(r_{\rm in})$ is obtained by integrating the PN
  equations starting from the initial conditions of the NR simulation,
  while the filled hexagon shows $\theta_{\rm fin}$ as computed by the
  NR simulation. \textit{Right:} The same as in the left panel, but for the configuration
  ``Q13TH00'' of Ref.~\cite{RITnew}. Clearly, only our formula reproduces the ``NR+PN''
  results to within 2-3 degrees for any $r_{\rm in}$, while the other formulas 
  have errors that grow very large for separations $r_{\rm in}\gtrsim 200M$.}
\label{fig:evolution}
  \end{center}
\end{figure}

To highlight the role played by the precession 
of the spins in the prediction of the final spin direction, in the left panel of Fig.~\ref{fig:evolution} we focus on
the binary ``SP3'' of Ref.~\cite{sp34} (index $41$ in Fig.~\ref{fig:combo}). In this configuration the spins precess
strongly and the final spin flips relative to $\boldsymbol{S}(r_{\rm
  in})$. In particular, $\boldsymbol{S}_1$ and $\boldsymbol{S}_2$ are
initially parallel, orthogonal to $\boldsymbol{L}$ and have:
$a_1=a_2\approx0.5$, $q=1$. The filled hexagon in
Fig.~\ref{fig:evolution} is the numerical result as obtained with an
initial separation $r_{\rm in} \simeq 6.6\,M$, while different lines
show the angle $\theta_{\rm fin}$ obtained with the various formulas
as a function of the binary separation. More specifically, using the
initial conditions of the NR simulation we have evolved the ``SP3''
binary back in time using the PN equations up to a separation of
$2\times10^4 M$, applying the various formulas at each step to compute
$\theta_{\rm fin,*}=\arccos[\boldsymbol{\hat{L}}(r_{\rm in}) \cdot
  \boldsymbol{\hat{a}}_{\rm fin,*}]$. The line ``PN+NR'' reports instead
the angle $\arccos[\boldsymbol{\hat{L}}(r_{\rm in}) \cdot
  \boldsymbol{\hat{a}}_{\rm fin, NR}]$. 
In the right panel, we consider the configuration 
``Q13TH00'' of Ref.~\cite{RITnew} (index $1$ in Fig.~\ref{fig:combo}). This configuration features
a smaller, non-spinning BH and a larger, spinning BH with spin $a\approx0.81$ lying on the equatorial 
plane and orthogonal to the separation ($r_{\rm in}\approx 6.4 M$). 
The mass ratio is $q\approx0.13$, which makes this binary interesting as
it allows to test the behavior of the various formulas in a region of parameter
space which so far has been sampled only sparsely by NR simulations, which usually deal with $q\approx1$. 
(This is because the cost of a NR simulation grows as $1/q^2$, a factor $1/q$ coming from the slow inspiral of a
small mass ratio binary, and another factor $1/q$ appearing because if the mass ratio is small, one needs much resolution 
to resolve the two very different scales of the
problem.) 

Three results are clear from
these comparisons. First: all previous formulas provide a reasonable
estimate of $\theta_{\rm fin}$ but only when the binary has a 
separation $r_{\rm in}\ll 200 M$. Second: 
Only our formula provides a reasonable
estimate at all separations; indeed, the ``PN+NR'' results are
reproduced to within the accuracy of the NR simulation, \textit{i.e.}~2-3 degrees. All
the other formulas, instead, predict the same value for $\theta_{\rm
  fin}$ for all $r_{\rm in}$,
 because they take as an input
the angles
 $\arccos({\boldsymbol{\hat{S}}}_{1,2}\cdot{\boldsymbol{\hat{L}}})$ and $\arccos({\boldsymbol{\hat{S}}}_{1}\cdot{\boldsymbol{\hat{S}}}_2)$, which
 are constant under the quasi-circular PN evolution~\cite{bcv}. Therefore, the other formulas
become rapidly become imprecise when the binary's separation increases, reaching errors as large as 
60 degrees for $r_{\rm in}\gtrsim200M$. Third: the error of the old formulas when applied to large separation
binaries is larger for small mass ratio systems (\textit{e.g.}~the maximum error is $\sim13$ degree for the comparable-mass
configuration SP3, while it is $\sim60$ degrees for the small mass-ratio binary Q13TH00). This had to
be expected, again on the grounds that the old formulas predict the same $\theta_{\rm fin}$ for all $r_{\rm in}$.
As a result of this, in fact, the maximum error of the old formulas is roughly given by their prediction for
 $\theta_{\rm fin}$ at small separations, because the correct $\theta_{\rm fin}$ becomes small for large $r_{\rm in}$ (\textit{cf.} the ``PN+NR'' curves in Fig.~\ref{fig:evolution}). This prediction can be very large for small $q$ if the angle between $\boldsymbol{\hat{S}}$ and $\boldsymbol{\hat{L}}$ at small separations is large. This can be easily understood by
noting that at small separations the old formulas give roughly the same results as our formula: Therefore, for $q\approx0$
their prediction is $\cos(\theta_{\rm fin}) \approx \boldsymbol{\hat{L}}(r_{\rm in}) \cdot \boldsymbol{\hat{S}}(r_{\rm in})$
[because for $q\approx0$,
$ \boldsymbol{\hat{J}}(r_{\rm in})\approx\boldsymbol{\hat{S}}(r_{\rm in})$].  

We stress that such a good agreement between our formula and the data for the final spin, 
irrespective of the separation of the
binary to which the formula is applied, 
emerges because we have consistently 
taken into account the effect of precession through assumptions
\textit{(iv)} and \textit{(v)}, while that effect is 
not accounted for in the other formulas. For instance, in our earlier formula of
Ref.~\cite{old} (``AEIold''), assumptions \textit{(iv)} and \textit{(v)} 
were replaced by the assumption that the angular momentum
emitted during the inspiral is parallel to $\boldsymbol{L}(r_{\rm in})$ 
[\textit{cf.} assumption \textit{(iii)} of Ref.~\cite{old}]. 
This is not true unless one neglects spin-orbit precession (as was indeed stressed explicitly in Ref.~\cite{old}). 
A similar assumption was also made in Ref.~\cite{bkl}, while Ref.~\cite{fau} admittedly recognized
that their formula might not be valid at arbitrary separations, as it is 
based on a fit to NR simulations, which, as already mentioned, have initial separations $r_{\rm in}\sim10 M$.

In conclusion: I have reviewed the assumptions needed to derive a new formula predicting the spin of
the BH resulting from the merger of two BHs in quasi-circular orbits
and having arbitrary initial masses and spins~\cite{new}. This formula includes the effect of the precession of the spins through
assumptions \textit{(iv)} and \textit{(v)}, and can therefore be applied to widely separated binaries, such as those relevant for cosmological
applications, for which the other available formulas become imprecise.
I stress that requiring that the formulas for the final spin work also when
applied to large separation binaries is necessary to ensure that these formulas 
can be readily used in cosmological applications. Cosmological simulations typically
provide the initial spins and angular momentum of SMBH binaries when their separation
is still of about 0.1 pc (see Ref.~\cite{marta} for a recent example), which corresponds to $2\times10^5 M$ if $M=10^8 M_\odot$. This
separation roughly corresponds to the point at which  
the evolution of the binary starts being driven by gravitational-wave emission.
While it would be in principle possible to use the PN equations and evolve the
widely-separated binaries provided by cosmological simulations to small separations, read-off the relevant
information and apply the formulas for the final spin, this procedure would be impractical 
and potentially very time-consuming (\textit{cf.} again Ref.~\cite{marta}, which simulates tens of thousands of SMBH binaries 
in order to get significant statistics).
 
\smallskip

% \section*{Acknowledgements}
\noindent I am grateful to my coauthor L. Rezzolla for countless 
discussions on the issues examined in this contribution. I also acknowledge support from NSF Grant PHY-0603762.


\section*{References}
\smallskip
\begin{thebibliography}{50}
\bibitem{new}  Barausse E and Rezzolla L 2009, \apj  {\bf 704} L40
\bibitem{bcv}   Buonanno  A, Chen Y and Vallisneri  M 2006, \prd {\bf 74} 029904 
\bibitem{scott} Hughes S~A and Blandford R~D\ 2003, \apjl  {\bf 585} L101
\bibitem{bkl} Buonanno A, Kidder L~E and Lehner L\ 2008, \prd  {\bf 77} 026004 
\bibitem{kesden} Kesden M\ 2008, \prd  {\bf 78} 084030
\bibitem{fit1}  Rezzolla L, Dorband E~N, Reisswig C, Diener P, Pollney D, Schnetter E and Szil{\'a}gyi B\ 2008, \apj  {\bf 679} 1422
\bibitem{fit2}  Rezzolla L, Diener P, Dorband E~N, Pollney D, Reisswig C, Schnetter E and Seiler J\ 2008, \apjl  {\bf 674} L29 
\bibitem{faunew} Tichy W and Marronetti P\ 2008, \prd {\bf 78} 081501
\bibitem{boyle1} Boyle L, Kesden M and Nissanke S\ 2008, \prl,  {\bf 100} 151101 
\bibitem{boyle2} Boyle L and Kesden M\ 2008, \prd  {\bf 78} 024017
\bibitem{old}  Rezzolla L, Barausse E, Dorband E~N, Pollney D, Reisswig C, Seiler J and Husa S\ 2008, \prd  {\bf 78} 044002 
\bibitem{caltech_cornell} Scheel M~A, Boyle M, Chu T, Kidder L~E, Matthews K~D and Pfeiffer H~P\ 2009, \prd  {\bf 79} 024003 
\bibitem{apostolatos}   Apostolatos T~A, Cutler C, Sussman G~J and Thorne K~S\ 1994, \prd  {\bf 49} 6274
\bibitem{RITnew} Lousto C~O and Zlochower Y\ 2009, \prd  {\bf 79} 064018 
\bibitem{sp34} Campanelli M, Lousto C~O, Zlochower Y, Krishnan B and Merritt D\ 2007, \prd  {\bf 75} 064030
\bibitem{sp6} Campanelli M, Lousto C, Zlochower Y and Merritt D\ 2007, \apjl  {\bf 659} L5 
\bibitem{fau} Tichy W and Marronetti P\ 2007, \prd {\bf 76} 061502
\bibitem{marta} Dotti  M, Volonteri  M, Perego  A, Colpi  M, Ruszkowski  M and Haardt  F\ 2010, \mnras {\bf 402} 682 

\end{thebibliography}
\end{document}